# Electrical contacts for high performance optoelectronic devices of BaZrS$_3$ single crystals

Huandong Chen[1,4], Shantanu Singh[1], Mythili Surendran[1,2], Boyang Zhao[1], Yan-Ting Wang[1], Jayakanth Ravichandran[1,2,3*]

[1]Mork Family of Department of Chemical Engineering and Materials Science, University of Southern California, Los Angeles CA, USA

[2]Core Center for Excellence in Nano Imaging, University of Southern California, Los Angeles, California, USA

[3]Ming Hsieh Department of Electrical and Computer Engineering, University of Southern California, Los Angeles CA, USA

[4]Present address: Condensed Matter Physics and Materials Science Department, Brookhaven National Laboratory, Upton, NY, USA

*Email: j.ravichandran@usc.edu

# Abstract

Chalcogenide perovskites such as $BaZrS_3$ are promising candidates for next generation optoelectronics such as photodetectors and solar cells. Compared to widely studied polycrystalline thin films, single crystals of $BaZrS_3$ with minimal extended and point defects, are ideal platform to study the material's intrinsic transport properties and to make first-generation optoelectronic devices. However, the surface dielectrics formed on $BaZrS_3$ single crystals due to sulfation or oxidation have led to significant challenges to achieving high quality electrical contacts, and hence, realizing the high-performance optoelectronic devices. Here, we report the development of electrical contact fabrication processes on $BaZrS_3$ single crystals, where various processes were employed to address the surface dielectric issue. Moreover, with optimized electrical contacts fabricated through dry etching, high-performance $BaZrS_3$ photoconductive devices with a low dark current of 0.1 nA at 10 V bias and a fast transient photoresponse with rise and decay time of < 0.2 s were demonstrated.

KEYWORDS: chalcogenide perovskite, single crystal, photoconductor, electrical contacts, $BaZrS_3$

# Introduction

Chalcogenide perovskites such as $BaZrS_3$ have recently emerged as one of the most promising material candidates for next generation photovoltaic applications due to their ideal optoelectronic properties[1-4], excellent stability[5,6] and earth abundancy[7,8]. Over the past few years, significant progress has been made on growing polycrystalline thin films of $BaZrS_3$ through processes such as sputtering[9], pulsed laser deposition[10,11], molecular beam epitaxy[12,13], and solution-based processing[14,15], while synthesizing high quality, large single crystals of $BaZrS_3$ remains challenging[16]. Unlike wafer-scale single crystalline Si or halide perovskites, the largest crystal sizes of $BaZrS_3$ one can obtain so far are only 100 – 200 μm (through the $BaCl_2$-flux growth method[16]), making it difficult to make any meaningful single crystal devices of $BaZrS_3$. On the other hand, it is of particular importance to do so, considering that compared with polycrystalline thin films, single-crystal-based optoelectronic devices are typically expected to exhibit the material's intrinsic transport properties and show better performance due to reduced defects and minimized grain boundaries, as has been validated in single crystalline Si[17,18] and halide perovskites solar cells[19-22].

Recently, a device fabrication strategy has been demonstrated to overcome such crystal size limitations on making electrical contacts, where the bulk crystal is embedded in a polymeric medium to flatten the top surface such that lithography-based contact fabrication and optimization procedures can be readily applied[23-25]. Here, we employ this strategy to micro-scale $BaZrS_3$ single crystals for electrical contact optimization and device fabrication. However, different from many other crystals that are electrically conductive and air-stable such as $BaTiS_3$, several important aspects need to be additionally considered when making electrical contacts on semiconducting $BaZrS_3$ crystals[26,27]: 1) sulfate or sulfite[26] layers are easily formed during the DI water wash-off step of the flux growth[27], making the surfaces of as-retrieved crystals insulating, and 2) the Zr species tend to be gradually oxidized in air or under heat treatment, as illustrated in Figure 1b.

In this work, we present the optimization of electrical contact fabrication processes on planarized sub-millimeter-sized single crystals of $BaZrS_3$ that addresses the issue of surface dielectrics. Al/$BaZrS_3$/Al-based photoconductive devices fabricated through gentle mechanical polishing and metal etch-back route showed μA level of dark current despite the obvious photo responses under illumination; While by employing polishing-free contact fabrication procedures through dry etching and electrodes lift-off, high-performance $BaZrS_3$ devices with a significantly reduced dark current level (0.1 nA at 10 V bias) and a fast photo response with a rise and decay time of < 0.2 s were achieved. Our work sheds light on developing other high-performance optoelectronic devices of single crystalline $BaZrS_3$ such as Schottky diodes and photovoltaics, as well as exploring its intrinsic transport properties.

**Existing issues for contact optimization of $BaZrS_3$ crystals**

Single crystals of $BaZrS_3$ with a space group of Pnma (Figure 1a) were synthesized using a molten $BaCl_2$ salt flux method[16]. Attempts have been made by the authors in early days to directly contact the as-grown $BaZrS_3$ surface using tungsten (W) probes, where no measurable current (< 10 pA) can be extracted up to 20 V bias from *I-V* measurements. One thing to notice is that DI water has been employed to wash off the excess salt flux after the growth for a prolonged time (~ 10 min), in order to retrieve the crystals[16,27]. Therefore, it is postulated that a layer of water-induced dielectric, either sulfate or sulfite[26] (Figure 1b), with certain thickness, has already been formed during the DI wash-off process for crystal retrieval, and hence, no electrical current could be extracted from the surface.

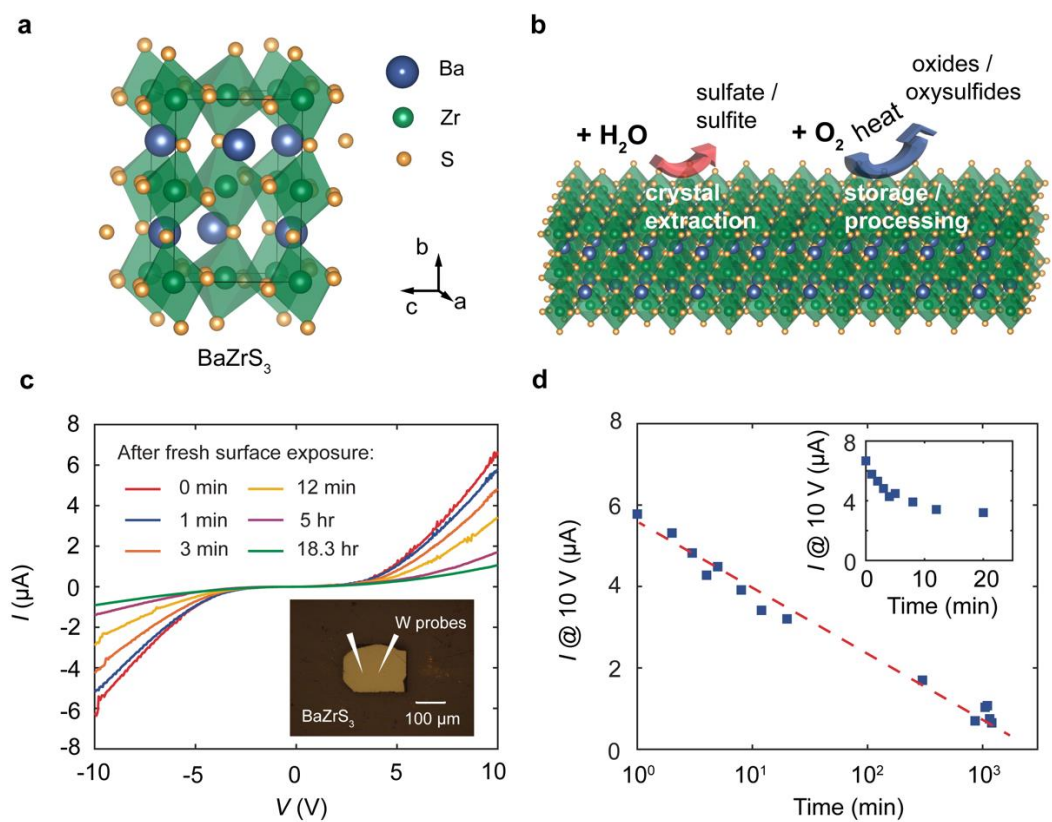

**Figure 1 Propensity of forming surface dielectrics on BaZrS$_3$.** (**a**) Schematic crystal structure of BaZrS$_3$. (**b**) Schematic illustration showing potential origins of surface dielectrics formation on BaZrS$_3$ bulk crystals. (**c**) *I-V* characteristics of a BaZrS$_3$ crystal as a function of time, where W probes were directly applied to contact the freshly exposed BaZrS$_3$ surface right after a gentle mechanical polishing. (**d**) Time dependence of the corresponding extracted dark current at 10 V, which follows a rough exponential decay as indicated by the dashed line. The inset shows the evolution of extracted current in the first 20 min in a linear scale.

Moreover, there is a propensity of surface oxide formation on BaZrS$_3$ crystal surfaces in air over prolonged time or upon heating (Figure 1b), similar to what has been reported on many other air-sensitive sulfides containing Zr species[28,29]. Here, we show complementary evidence from electrical measurements. Time-dependent *I-V* measurements were carried out on freshly exposed BaZrS$_3$ surface (through mechanical polishing) using W probes, as illustrated in Figure 1c and Figure 1d. Right after gentle hand-polishing process, up to 6 µA current was extracted at 10 V without illumination, although the *I-V* characteristics were not linear. A sharp drop of current was observed within the first 10 min of surface exposure and it was gradually reduced to close to 1 µA after about 60 hr. One potential explanation is that a thin layer of oxide has been formed over time on the BaZrS$_3$ fresh surface, leading to largely modulated *I-V* characteristics when using W probes. Recently, a scanning transmission electron microscopy (STEM) study on BaZrS$_3$ single crystal prepared by focused ion beam (FIB) lift-out has clearly shown a few nm of surface amorphous oxides on the surface, which is presumably formed during the TEM sample preparation or storage procedures[27], consistent with our observations. Therefore, care needs to be taken to freshly expose BaZrS$_3$ surface right before electrode fabrication for obtaining high-quality electrical contacts.

Moreover, to overcome the dimension-induced challenges in handling BaZrS$_3$ crystals, all the BaZrS$_3$ device fabrication processes presented in this work started by embedding a micro-scale BaZrS$_3$ single crystal in a polymeric medium. The detailed embedding processes are modified from previous reports[23,30,31] and are illustrated in Figure S1. Such procedures lead to a planarized BaZrS$_3$ crystal top surface that is suitable for standard cleanroom fabrication processes such as lithography and etching, which has enabled all the contact optimization processes for BaZrS$_3$ single crystal devices that are discussed in this manuscript.

**Device fabrication through mechanical polishing and aluminum etch-back**

We first applied a gentle mechanical polishing step to get rid of all the water-induced surface dielectrics from an as-embedded BaZrS$_3$ crystal. To minimize the formation of surface oxides during fabrication processes upon heating and prolonged exposure to air, the sample was loaded into the vacuum chamber for sputtering of 200 nm aluminum (Al) film right after polishing. Regular photolithography was then employed on Al film to form a two-terminal pattern, followed by a wet chemical etching procedure of aluminum in KOH solution (0.1 mol/L) to form the electrodes. Figure 2a shows representative optical microscopic images of a BaZrS$_3$ device showing various device fabrication processes after polymeric planarization (left), mechanical polishing and Al deposition (middle), and Al etch-back step (right). Importantly, no obvious chemical attacks or damages of BaZrS$_3$ were observed during KOH etching, which has enabled the fabrication of Al/BaZrS$_3$/Al-based devices through wet etching. Alternatively, one may also seek to fabricate electrodes on BaZrS$_3$ single crystals using regular photolithography and metal lift-off after crystal polishing[27], which features a much wider range of metal selection for contact optimization. However, in that scenario, the chances of introducing surface oxides with direct lift-off route can be higher, considering that freshly exposed BaZrS$_3$ crystals need to go through several heat treatment processes (~ 110°C) and potential exposure to air during the lithography procedures.

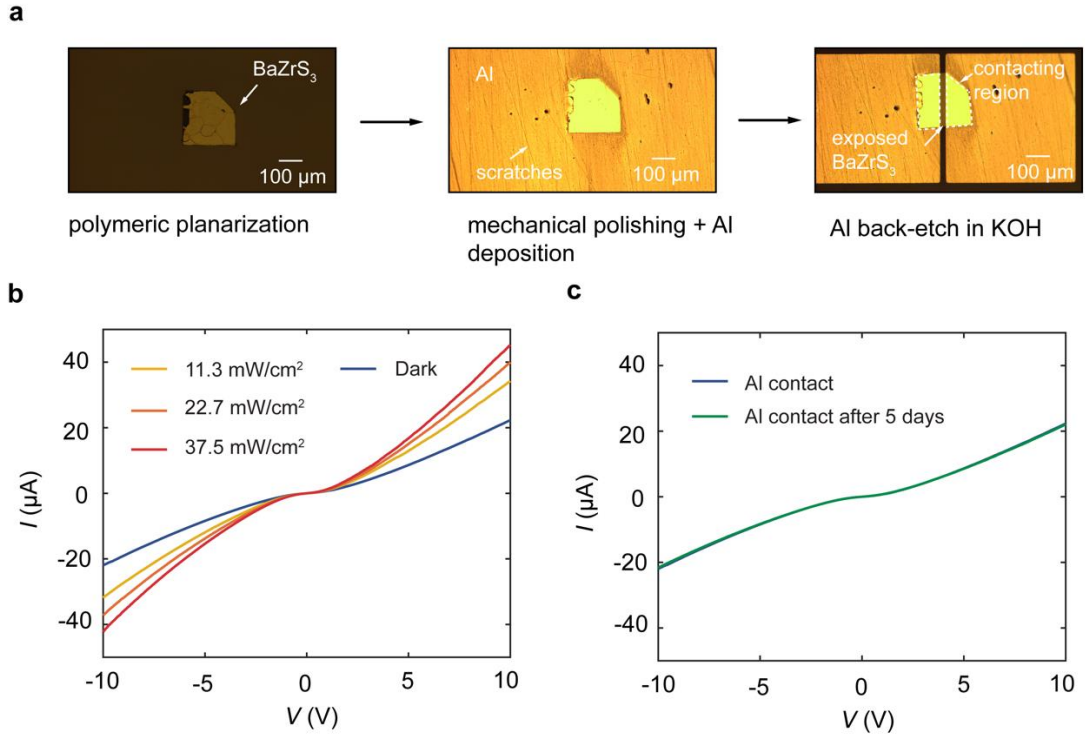

**Figure 2 Al/BaZrS₃/Al bulk photoconductive device fabrication and characterization.** (**a**) Process flow of a two-terminal Al/BaZrS$_3$/Al photoconductive device with optical microscopic images of a BaZrS$_3$ crystal after polymeric planarization (left), mechanical polishing and Al deposition (middle), and Al electrodes patterning (right). (**b**) Representative *I-V* characteristics of an Al/BaZrS$_3$/Al device under different illumination, indicating obvious photoconductive responses of BaZrS$_3$. (**c**) Dark *I-V* characteristics of the same Al/BaZrS$_3$/Al device after 5 days.

Figure 2b shows representative *I-V* characteristics of such an Al/BaZrS$_3$/Al photoconductive device without illumination and under three different illumination intensities. Nonlinear *I-V* curves were observed when directly using sputtered Al as the contact, however, a direct connection between the work function of Al and the band edge of BaZrS$_3$ is yet hard to make, considering the involvement of polishing-induced defect states that complicate the band alignment at the Al/BaZrS$_3$ interface. Nonetheless, a clear photo response was clearly resolved with a dark current of about 21 μA at 10 V and a current of 43 μA extracted at 37.5 mW/cm$^2$ broadband illumination (~ 0.4 suns). We also examined the reliability of Al contacts by measuring its *I-V* characteristics after 5 days, as shown in Figure 2c, which is in stark contrast to the measurements performed directly using W probes.

On the other hand, it is also noted that the Al/BaZrS$_3$/Al device features tens of μA levels of dark current at 10 V external bias. Such large values of dark current are typically associated with substantial amount of point defects such as sulfur vacancies, as has been reported in BaZrS$_3$ thin films processed at high temperatures[6,32]. In this experimental scheme, the most probable source of defects in BaZrS$_3$ is thought to be the mechanical polishing process, as evidenced by the recent observation of extended defects (such as dislocations, stacking faults, and grain boundaries) populated down to ~ 300 nm beneath the surface in mechanically polished BaZrS$_3$ crystal using electron microscopy[27]. Recently, In/Ga eutectic was attempted to contact a mechanically cleaved BaZrS$_3$ crystal directly without polishing for a proof-of-concept demonstration, where promising device performance with linear *I-V* responses and a low dark current of several nA at 10 V external bias were demonstrated[27]. However, such electrical contacting method is not ideal for practical device preparation and questions were also raised on the non-negligible amounts of defects formed during cleavage that would be detrimental to device performances. Therefore, developing lithography-compatible and polishing-free device fabrication processes for BaZrS$_3$ is highly demanded.

**Developing polishing-free fabrication processes for BaZrS$_3$**

Developing an etching route that effectively removes surface dielectrics and freshly exposes the underlying material in the contacting regions is essential for achieving high quality electrical contacts on bulk BaZrS$_3$ crystals. However, despite its importance and urgency, as a relatively new material system, neither of the wet etching or dry etching processes of BaZrS$_3$ or its surface dielectrics has been well studied in the literature.

We first present the wet etching route that we have attempted. The chemical etchant we used for BaZrS$_3$ is a mixture of buffered HF (BHF) solution and HCl (BHF : HCl : DI = 1 : 10 : 40, in volume, Recipe 1), which was originally developed and optimized for patterning 0.5Ba(Zr$_{0.2}$Ti$_{0.8}$)O$_3$-0.5Ba(Zr$_{0.7}$Ti$_{0.3}$)O$_3$ (BCZT) thin film[33], a relaxor ferroelectric that shares chemical elements with BaZrS$_3$. Here, both the original etchant and a DI-diluted version (BCZT etchant : DI = 1 : 10, in volume, Recipe 2) were tested for BaZrS$_3$ wet etching. Figure 3a shows optical microscopic images of a BaZrS$_3$ crystal (attached to a poly (dimethylsiloxane) PDMS stamp) before and after wet etching for a short period of time. It is clear that fresh BaZrS$_3$ surface was exposed after just a few seconds' etching (Figure 3a top right and bottom right). However, neither of the two etching recipes provides a good selectivity between the surface dielectrics and BaZrS$_3$ and their etching rates are too high (~ 20-30 µm/min for Recipe 1) for realizing well-controlled etching; An optimal etching time in the range of a few seconds using the diluted etchant (Recipe 2) is expected to effectively expose the fresh BaZrS$_3$ without introducing too much etching-induced damages, but the accurate controlling of this etch time is still rather challenging. Nonetheless, we found it very useful for shaping BaZrS$_3$ thin films devices. Figure S2 shows a good example of applying this wet chemical etchant (diluted version) for patterning pulsed laser deposition (PLD)-grown BaZrS$_3$ films[10], where the surface oxide issue is not as severe for fabricating electrical contacts and the growth substrate LaAlO$_3$ serves as a good etching stop, although the lateral etching can still become an issue due to the large etch rate. By further diluting the current chemical etchant for BaZrS$_3$, one may be able to obtain an ideal etchant with slow and controllable etching, which will serve well for fabricating various BaZrS$_3$ devices in the future regardless of its form.

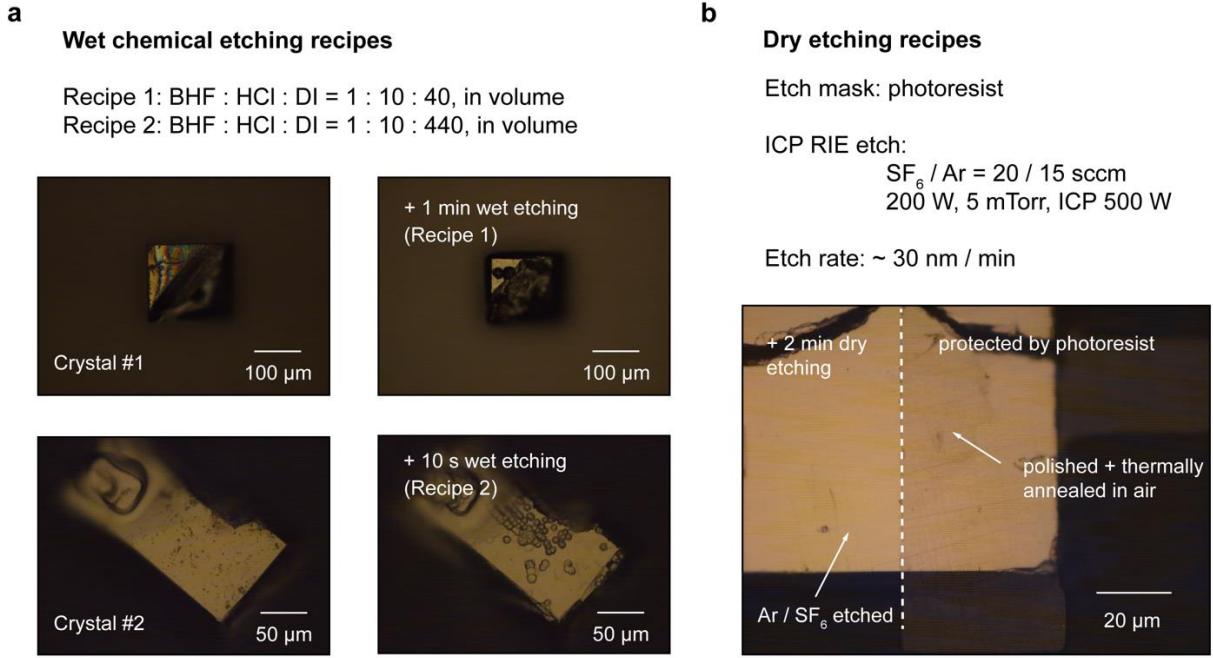

**Figure 3 Development of wet and dry etching recipes for BaZrS$_3$.** (**a**) Two different wet chemical etching recipes (HF- and HCl-based) tested for BaZrS$_3$. The bottom panel shows optical micrographic images of BaZrS$_3$ crystals before and after wet etching. (**b**) Ar/SF$_6$-based ICP-RIE dry etching recipe developed for BaZrS$_3$. The optical image below clearly illustrates the surface of a BaZrS$_3$ with (left, exposed to etching gases) and without (right, protected by photoresist) going through dry etching.

Further, we developed the dry etching route for BaZrS$_3$ and its surface dielectrics. Inductively coupled plasma – reactive ion etching (ICP-RIE) was employed using Ar and SF$_6$ as the etching gases. To test our dry etching recipes, a planarized BaZrS$_3$ crystal was first mechanically polished and thermally annealed in air (170˚C, 10 min) to form both scratches (used as marks) and surface oxides. The crystal was then patterned using a regular photoresist as the etching mask and dry-etched for 2 min (SF$_6$ 20 sccm, Ar 15 sccm, 200 W, 5 mTorr, ICP 500 W), as shown in Figure 3b. The etching rate of ~ 30 nm / min was determined by measuring the formed step height after dry etching. A relatively thick positive photoresist (~ 5-6 µm) was intentionally selected for now to survive the harsh dry etching process, considering a much larger etch rate of photoresist of ~ 1 µm / min under the same conditions. Further optimization on the dry etching recipes such as etching gases, pressure and power, or the employment of an etch mask with a much smaller etching rate towards the realization of an improved etching selectivity would unambitiously simplify the fabrication processes for BaZrS$_3$ devices.

## High-performance BaZrS$_3$ photodetectors fabricated through dry etching

For actual BaZrS$_3$ single crystal devices presented in this work, a two-step fabrication process was adopted in addition to the polymeric embedding procedure to planarize the crystal top surface: 1) electrical contacting areas were first defined by patterning vertical interconnect access (VIA) holes, followed by a dry etching step of 2-3 min to exposure fresh BaZrS$_3$ surfaces, as indicated in Figure 4a by the dashed white lines, and (2) a second photolithography procedure was applied for fabricating desired electrodes, with a short cleaning step (dry etching of ~ 30 s) carried out right before metal deposition. Therefore, the BaZrS$_3$ channel region remains intact throughout the fabrication processes. Figure 4a shows an optical microscopic image of such a two-terminal

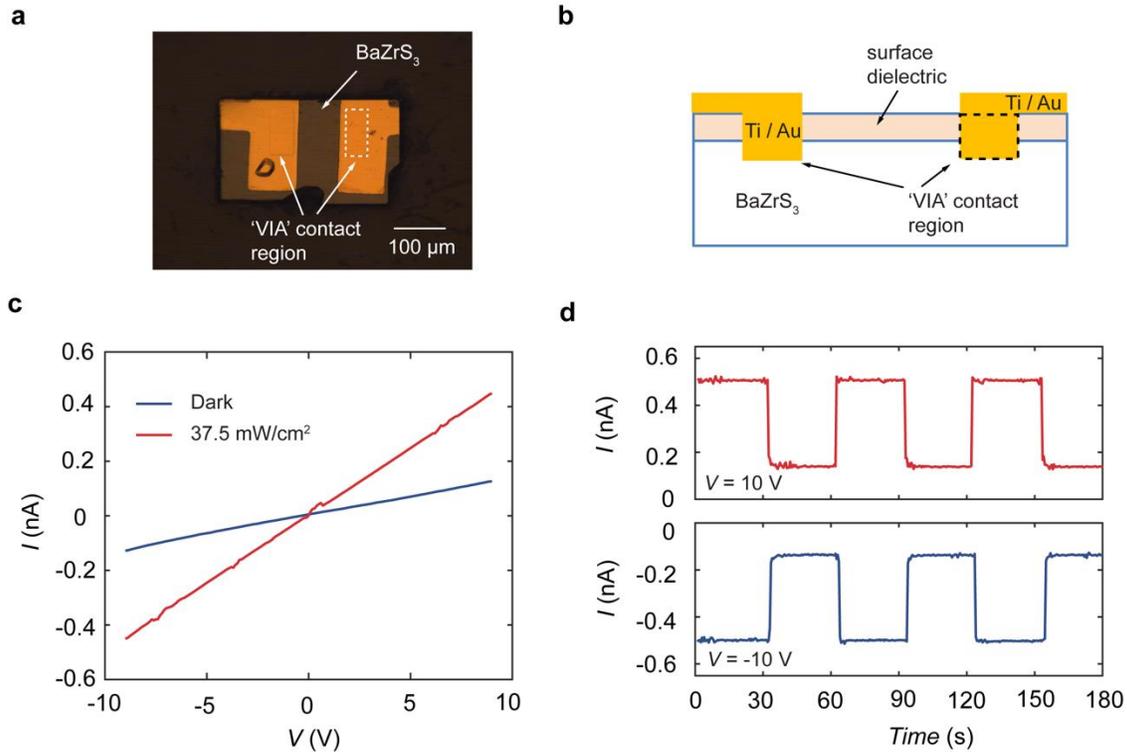

**Figure 4 Bulk BaZrS$_3$ photoconductive device fabricated through dry-etching and their characterization.** (**a**) Optical micrographic image of a two-terminal Au-BaZrS$_3$-Au photoconductive device fabricated through dry etching and metal lift-off processes. Dashed lines indicate the pre-defined 'VIA' contact region by dry etching. (**b**) Schematic illustration of a cross-sectional view of the BaZrS$_3$ device. (**c**) Representative *I-V* characteristics of Au-BaZrS$_3$-Au photoconductive devices under dark and illuminated conditions. (**d**) Transient photoresponse of the Au-BaZrS$_3$-Au photoconductive device at + 10 V and -10 V bias and white light illumination.

BaZrS$_3$ photoconductive device, and the cross-sectional view of the device is illustrated in Figure 4b.

We then carried out *I-V* characterization of single crystal photodetectors fabricated through polishing-free dry etching processes under different illumination intensities, as shown in Figure 4c. It is important to note that its dark current is only ~ 0.1 nA at 10 V applied bias, which is more than three order smaller than that of a BaZrS$_3$ single crystal device fabricated through mechanical polishing (Figure 2a and Ref[27]), two orders smaller than that of a PLD-grown BaZrS$_3$ thin film devices[10], and one order smaller than that of a mechanically cleaved BaZrS$_3$ single crystal device[27].

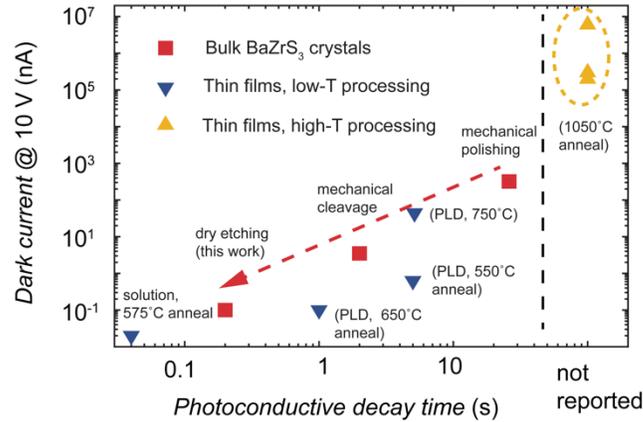

**Figure 5 Comparison of dark current and photoconductive decay time of previously reported BaZrS$_3$ photoconductive devices using different synthesize and fabrication routes.**

Such low levels of dark current are attributed to the small numbers of defects in the active channel region as it is fully protected throughout the entire device fabrication processes. Similar levels of dark current in BaZrS$_3$ have only been observed in low-temperature sulfur-annealed thin film devices, where sulfur vacancies are minimized[11,34]. Figure 4d shows the transient photo response of the same BaZrS$_3$ device measured at + 10 V and -10 V bias and under 37.5 mW/cm$^2$ illumination. The characteristic time constants of rise time and decay time were determined to be 0.18 s and 0.2 s, respectively, from a similar measurement using higher temporal resolution (Figure S3). Table S1 lists the performance of reported BaZrS$_3$ photoconductive devices using different synthesis and fabrication routes, and our devices show improved performance in terms of reduced dark current and faster response when comparing with any other BaZrS$_3$ single crystal devices[27] and most of BaZrS$_3$ thin film devices[6,10,11,32]. As illustrated in Figure 5, lowering synthesis and sulfur-annealing temperatures helps improve the performance of thin film BaZrS$_3$ photoconductive devices by reducing sulfur vacancies-based defects; while for single crystal BaZrS$_3$ devices, we achieved similar enhanced device performance by optimizing contact fabrication procedures. The dry-etching-based method introduces minimal amounts of defects during device fabrication processes on BaZrS$_3$ crystals and therefore, it helps preserve the intrinsic transport properties of the material.

## Conclusion and discussion

In conclusion, we have optimized contact fabrication processes to realize photoconductive devices of BaZrS$_3$ single crystals featuring low dark current and fast photo response. Our study paves the way for fabricating high performance optoelectronic devices of BaZrS$_3$ and sheds light on probing its intrinsic transport properties.

It is important to note that optoelectronic performance of BaZrS$_3$ seems to be susceptible to defects, as has been reported in this manuscript and many other previous studies[11,27]. Therefore, it is critical to minimize the generation of defects during device fabrication procedures, as we have demonstrated in this work. Alternatively, one may also apply sulfur annealing procedures to heal the surface oxides formed during flux wash-off step or seek to a different crystal method such as chemical vapor transport method, where the sulfur-rich growth environment shall minimize the surface oxide issues[35]. Moreover, inspired by the halide perovskites[36-39], the development of novel synthesis methods that lead to large and thin single crystals of BaZrS$_3$ and damage-free device fabrication procedures would greatly facilitate the realization of high-efficient BaZrS$_3$ single crystal-based photovoltaics.

## Methods

**Crystal growth and polymeric planarization**

Single crystals of BaZrS$_3$ were synthesized using a BaCl$_2$ flux growth method, as reported elsewhere[16]. DI water was used to wash off the excess flux for ~10 min to retrieve the crystals, which are typically in cubic morphologies and with a lateral dimension of 100-200 μm.

An as-grown micro-scale BaZrS$_3$ crystals with a thickness of ~ 100 μm and at least one flat surface was first attached to a poly PDMS stamp and emerged into a UV-curable epoxy (NOA 61, Norland Products, Inc) on a sapphire substrate (5 mm × 5 mm). The leveling of the PDMS surface was carefully adjusted in parallel to the substrate surface using a home-built transfer stage[40], followed by a UV-exposure step (30-min UV exposure step (B-100 *i*-line UV lamp, UVP)) to fully cure the epoxy before releasing the PDMS stamp. The planarized BaZrS$_3$ crystals were then used for different condition tests and actual device fabrication.

**Surface dielectric removal and device fabrication**

Electrical contacts were made on planarized BaZrS$_3$ crystals through lithography-based processes after removing the surface oxides, either by gentle mechanical polishing or wet / dry chemical etching.

For BaZrS$_3$ devices with back-etched Al contacts, we gently polished the crystal surface using a fine sanding paper (#12000, Micro-Mesh) right before the deposition of 200 nm Al thin film by sputtering. The Al electrodes were formed using regular photolithography and wet chemical etching of Al in 0.1 mol/L KOH solution for ~ 5 min.

To fabricate BaZrS$_3$ devices through dry etching, a first photolithography step was employed to define the contacting area using a thick photoresist (AZ 4620, MiroChemicals, 5-6 μm thick) as the etching mask, where BaZrS$_3$ surfaces were freshly exposed through an ICP-RIE etching step (SF$_6$ 20 sccm, Ar 15 sccm, 200 W, 5 mTorr, ICP 500 W) for 2-3 min. A second photolithography step was used to define electrical contacts (Ti/Au = 5/200 nm by Ebeam evaporation). A negative photoresist (AZ nLoF 2070, MicroChemicals, ~ 5.5 μm) was selected for the ease of lift-off, and a gentle surface cleaning process (SF$_6$ 20 sccm, Ar 15 sccm, 200 W, 5 mTorr, ICP 500 W, 30 s, or SF$_6$ 15 sccm, Ar 50 sccm, 100 W, 50 mTorr, ICP 500 W, 1.5 min) was carried out right before metal deposition.

The wet chemical etchant for both BaZrS$_3$ and its surfaces oxides was prepared by diluting a prepared mixture of BHF and HCl (BHF : HCl : DI = 1 : 10 : 40) etchant in DI water (pre-mixed etchant : DI = 1:10, in volume). The pre-mixture was prepared by first mixing HCl (36.5-38 %, VWR Chemicals BDH) and DI with a 1:4 ratio in volume, and then 1 part of BHF (7 : 1, J.T.Baker) was added to 10 parts of the diluted HCl mixture in volume. The fabrication details and performance of BaZrS$_3$ single crystal devices using wet chemical etching have not been extensively explored and are not discussed in this manuscript.

**Electrical characterization**

All electrical measurements were carried out on a probe station (M&M Micromanipulator) using a semiconductor analyzer (Agilent 4156C). Broadband illumination was applied on BaZrS$_3$ devices through the optical microscope of the probe station and transient photo responses (Figure 4d) were measured using the "sampling measurements" mode of the semiconductor analyzer. For the transient photoresponse presented in Figure S3, a white-light LED chip was used as the light source to generate a modulated broadband illumination with a 5 s switching period, which was powered by a function generator (SRS DS 340) with 0.2 Hz square wave output.

# Author declarations

**Author contributions**

H.C. and J.R. conceived the idea. H.C. designed the methodology and experiments. S.S. and B.Z. grew the crystals. H.C. optimized the fabrication processes and prepared devices. H.C. and S.S. carried out electrical characterizations, with input from M.S. and Y.W. H.C. and J.R. wrote the manuscript with input from all other authors.

**Acknowledgements**

The authors gratefully acknowledge the use of facilities at John O'Brien Nanofabrication Laboratory and Core Center for Excellence in Nano Imaging at University of Southern California for the results reported in this manuscript.

**Funding**

This work was supported by the Army Research Office under award numbers W911NF-21-1-0327 (MURI) and W911NF-24-1-0164, and the U.S. National Science Foundation under grant numbers DMR- 2122071. The crystal growth efforts were supported by the Office of Naval Research under grant number N00014-23-1-2818.

**Data availability**

The data are available from the corresponding author of the article on reasonable request.

**Conflict of interest**

The authors declare no competing financial interests.

**Graphic abstract**

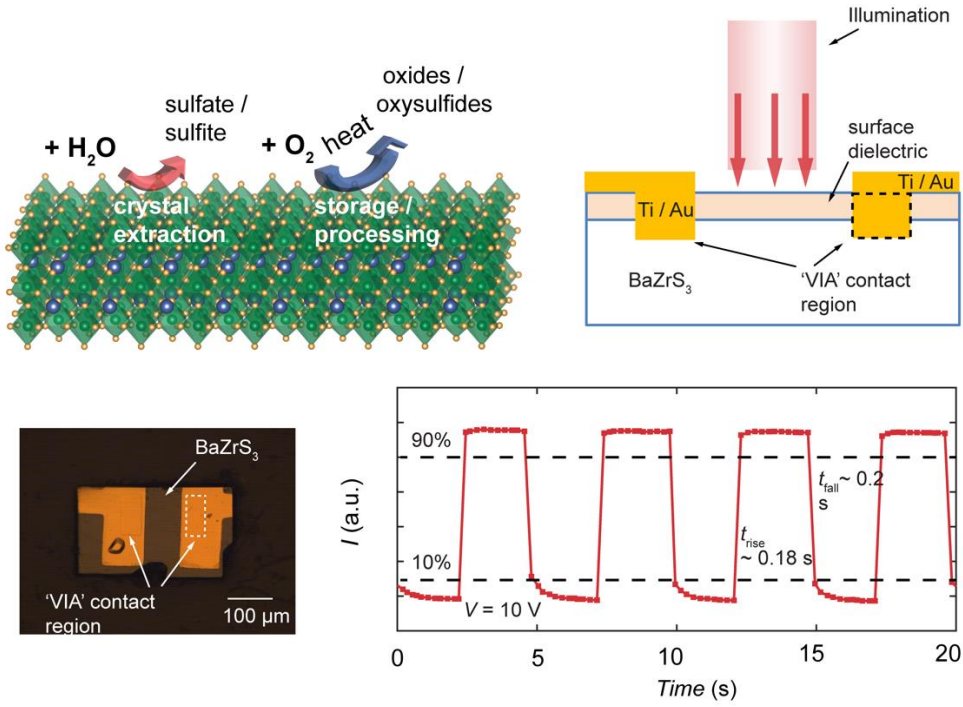

# Supplemental Information for

# Electrical contacts for high performance optoelectronic devices of BaZrS$_3$ single crystals

Huandong Chen[1,4], Shantanu Singh[1], Mythili Surendran[1,2], Boyang Zhao[1], Yan-Ting Wang[1], Jayakanth Ravichandran[1,2,3]*

[1]Mork Family of Department of Chemical Engineering and Materials Science, University of Southern California, Los Angeles CA, USA

[2]Core Center for Excellence in Nano Imaging, University of Southern California, Los Angeles, California, USA

[3]Ming Hsieh Department of Electrical and Computer Engineering, University of Southern California, Los Angeles CA, USA

[4]Present address: Condensed Matter Physics and Materials Science Department, Brookhaven National Laboratory, Upton, NY, USA

*Email: j.ravichandran@usc.edu

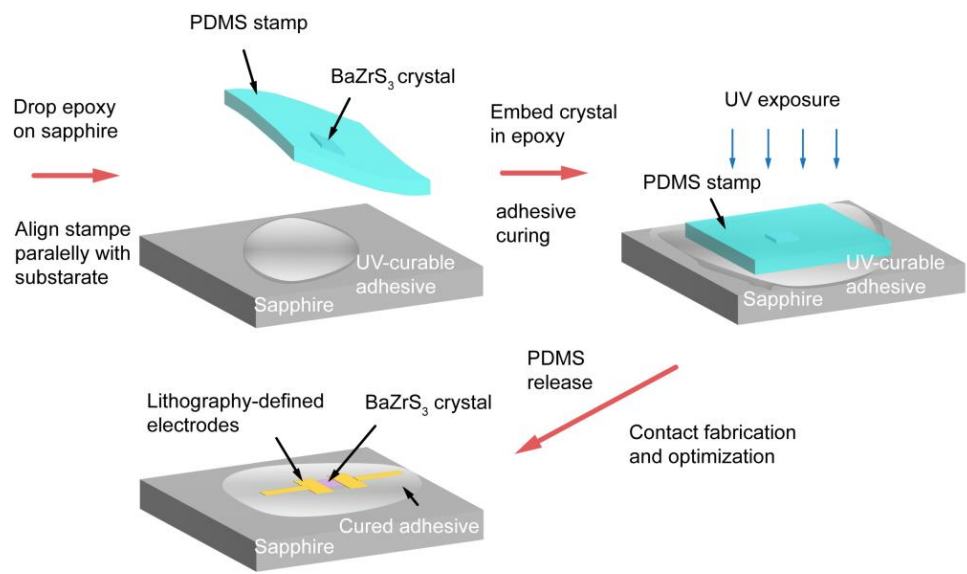

**Figure S1.** Schematics of the fabrication process flow for a bulk micro-scaled BaZrS$_3$ device utilizing a UV-curable adhesive as the embedding medium. All the contact fabrication and optimization procedures of bulk BaZrS$_3$ discussed in this manuscript started from an embedded BaZrS$_3$ crystal whose top surface has been flattened using the above processes.

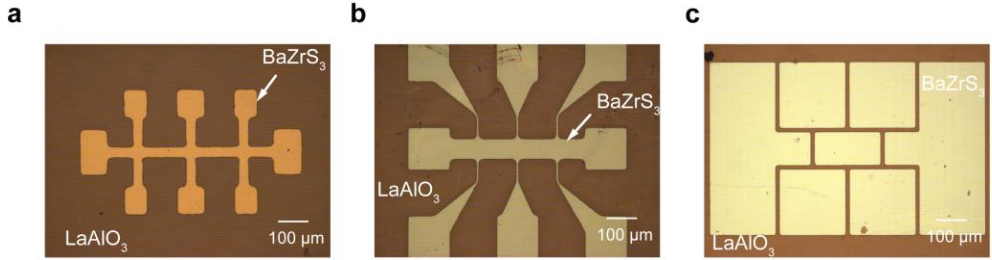

**Figure S2.** (**a**) to (**c**) Optical micrographic images of various patterned Hall bar devices using PLD-grown BaZrS$_3$ thin films. All the Hall bar devices were fabricated through photolithography and wet chemical etching of BaZrS$_3$ using the wet chemical etching Recipe 2 (BHF : HCl : DI = 1 : 10 : 440, in volume) as discussed in the manuscript.

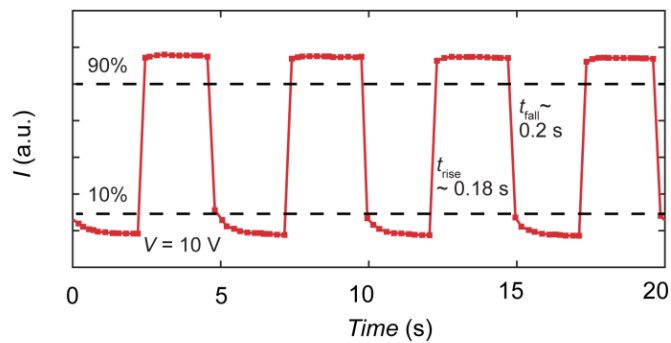

**Figure S3.** Transient photoresponse of the Au-BaZrS$_3$-Au photoconductive device measured at + 10 V bias. A white-light LED chip was used as the light source, which was powered by a function generator (SRS DS 340) with 0.2 Hz square wave output. The rise time $\tau_{rise}$ (time required to reach 90% of photocurrent after the illumination is switched on) and decay time $\tau_{decay}$ (time required to reach 10% of photocurrent after the illumination is switched off) were determined to be 0.18 s and 0.2 s, respectively.

**Table S1. Performance of reported BaZrS$_3$ photoconductive devices.**

| BaZrS$_3$ form | Synthesis route | Post-growth processing | Electrodes | Dark current | Illumination | $\tau_{rise}$ | $\tau_{decay}$ | References |
|---|---|---|---|---|---|---|---|---|
| Thin film | PLD-BaZrO$_3$ film sulfurization | Sulfurization at 1050°C, 4 hr | Au | 0.3 mA at 10 V | N/A | N/A | N/A | 1 |
| | | Sulfurization at 1050°C, 2 hr | Au | ~ 0.2 μA at 10 V | N/A | N/A | N/A | 1 |
| Thin film | BaZrO$_3$ film (chemical solution deposition) sulfurization | Sulfurization at 1050°C, 4 hr | Au | 3 mA at 5 V | N/A | N/A | N/A | 2 |
| Thin film | PLD | none | Ti/Au | 22 nA at 5 V | 532 nm laser | 0.72 s | 5.14 s | 3 |
| Thin film | PLD | Post-annealing at 650°C | Au | ~ 0.1 nA at 10 V | 532 nm laser | < 0.5 s* | ~ 1 s* | 4 |
| | | Post-annealing at 550°C | Au | 0.5 nA at 8 V | 532 nm laser | ~ 2 s | ~ 5 s | 4 |
| Thin film | Solution process | Post-annealing at 575°C | Au | ~ 0.01 nA at 5 V | 630 nm laser | ~ 40 ms | ~ 40 ms | 5 |
| Single crystal | BaCl$_2$ flux | Mechanical polishing | Ti/Au | 320 nA at 10 V | White light | 18 s | 26 s | 6 |
| | | Mechanical cleavage | In/Ga eutectic | 3.5 nA at 10 V | White light | 2 s | 2 s | 6 |
| Single crystal | BaCl$_2$ flux | Dry etching | Ti/Au | 0.1 nA at 10 V | White light | 0.18 s | 0.2 s | This work |

**\*** The rise and decay time is extracted from the digitized data of the original curve, and the transient time may be overestimated due to the lack of a zoomed-in plot.